\documentstyle[preprint,aps]{revtex}

\begin{document}

\draft

\title{LOWER BOUND ON THE PION POLARIZABILITY\\
FROM QCD SUM RULES}
\author{M. Benmerrouche$^{ a}$, G. Orlandini$^{ b}$
and T.G. Steele$^{ a}$}
\address{$^{ a}$Saskatchewan Accelerator Laboratory,\\
University of Saskatchewan, Saskatoon, Saskatchewan S7N 0W0, Canada\\
$^{ b}$Dipartimento di Fisica, Universit\`a di Trento,\\ I-38050 Povo
(Trento) Italy}

\date{\today}

\maketitle

\begin{abstract}

\noindent Making use of QCD sum rules a lower bound
is found which
relates the electromagnetic polarizability
$\alpha_{_E}$ and mean-square radius $\langle r_\pi^2\rangle$
of charged pions through the intrinsic polarizability
$\tilde\alpha_{_E}=
\alpha_{_E}-\alpha\langle r_\pi^2\rangle/(3M_\pi)$.
 We find that if present constraints on the QCD
continuum (duality) threshold are accepted, this lower bound on the
intrinsic polarizability
$\tilde\alpha_{_E}$
is incompatible with
some previous determinations of $\alpha_{_E}$ and $\langle r_\pi^2\rangle$.

\end{abstract}

\vspace{0.5in}

Recently the electromagnetic polarizability
of charged pions has been the
object of several theoretical and experimental investigations, since it
is believed that one can make direct predictions based upon
quantum chromodynamics (QCD).

In the standard unit of $10^{-4}\,{\rm fm^3}$ (which will be used throughout
this paper), chiral perturbation theory predicts $\alpha_{_E}=2.8$
\cite{Holstein90,DonoghueHolstein93}, while
alternative approaches based on the Das-Mathur-Okubo (DMO) sum rule
\cite{DasMathurOkubo67}
have resulted in values ranging from $2.6$ (based on resonance
saturation\cite{Holstein90}) to $5.6\pm\,0.5$\cite{Lavelleetal94}.
Other theoretical calculations in quark and NJL models range between
$3.6$ and $13$\cite{othertheory}. On the experimental side there
are three measurements of $\alpha_{_E}$, ranging from
$2.2\pm\,1.6$\cite{Babuscietal92} to $6.8\pm\,1.4\,\pm 1.2$
\cite{Antipovetal85}, and to $20\pm\,12$\cite{Aibergenovetal86}.

In view of the large experimental and theoretical uncertainties it is
worthwhile to search for theoretical bounds related to  $\alpha_{_E}$.
This is done in this work using the relation between the
polarizability and the axial current form factor in the decay
 $\pi\to\,l\, \nu\,\gamma$ \cite{terentiev73} and
the above mentioned current
algebra sum rule
\cite{DasMathurOkubo67}
which states that the intrinsic polarizability $\tilde\alpha_{_E}$
is given by
\begin{equation}
\tilde\alpha_{_E}=\alpha_{_E}-{\alpha\over 3 M_\pi} <r^2_{\pi}> = {\alpha\over
M_{\pi}\,f^2_{\pi}}\,\int\limits
_{4 M^2_{\pi}}^\infty dt {1\over t}\,[\rho^A(t)-\rho^V(t)]\,,
\label{dmo}
\end{equation}
where $M_{\pi}$ is the pion mass, $\langle r_{\pi}^2\rangle$ is the mean-square
pion radius, and $f_{\pi}=133\,$MeV is the pion decay constant.
In the derivation of the DMO sum-rule, the pion pole contribution is explicitly
removed from the right-hand side in (\ref{dmo}) as reflected by the
$4\,M_\pi^2$
threshold. The value of the mean-square radius is also uncertain, with
values ranging from $\langle r_\pi^2\rangle=6/M_\rho^2=0.4\,{\rm fm}^2$ from
vector meson dominance, to the experimental results
$\langle r_\pi^2\rangle = (0.439\pm 0.008)\,{\rm fm^2}$
\cite{amend} and $\langle r_\pi^2\rangle=(0.463\pm 0.006)\,{\rm fm^2}$
\cite{gesh}.

The spectral functions $\rho^A(t)$ and $\rho^V(t)$
are given by the absorptive parts of the axial and vector current correlators,
respectively:

\begin{equation}
\rho^{A,V}(t) = {1\over\pi} Im\,\Pi^{A,V}(q^2)\,,
\label{spectral}
\end{equation}
where
\begin{equation}
\Pi^{A,V}(q^2)={4\over3} \,i
\left({q_{\mu}q_{\nu}\over q^2}-{1\over4}g_{\mu\nu}
\right) \int d^4x e^{iq\cdot x}<0|T(J^{A,V}_\mu(x)J^{A,V}_\nu(0))|0>\,,
\label{corr}
\end{equation}
with
\begin{equation}
J^V_\mu(x)=\bar u\gamma_\mu d\,,\,\,\,\,\,\,\,\,\,\,\,\,\,\,\
J^A_\mu(x)=\bar u\gamma_\mu \gamma_5 d\quad .
\label{curr}
\end{equation}

Of course large differences in the value of the intrinsic polarizability
$\tilde\alpha_{_E}$ arise because
the sum rule involves cancellation of large numbers between the vector and
axial
spectral functions. To avoid this problem,
we will show that for each term of the sum rule one can
find interesting upper and/or lower bounds based on QCD sum rules.
These bounds are a function QCD condensates and of
the continuum (duality)
thresholds.
A suitable combination of these bounds results in a meaningful lower
bound on the intrinsic polarizability $\tilde\alpha_{_E}$ relating
$\alpha_{_E}$ and $\langle r_\pi^2\rangle$.

Our main aim is to find a lower bound to the integral in (\ref{dmo})
\begin{equation}
I=\int\limits_{4 M^2_{\pi}}^\infty dt {1\over t}\,[\rho^A(t)-\rho^V(t)]\,.
\label{I}
\end{equation}
This can be rewritten as
\begin{equation}
I=I^A\left(s_0^A\right)-I^V\left(s_0^V\right)+
C^{AV}\left(s_0^A,s_0^V\right)\,,
\label{I2}
\end{equation}
where
\begin{eqnarray}
I^{A,V}\left(s_0^{A,V}\right)&=&\int\limits_{4 M^2_{\pi}}^{s_0^{A,V}} dt
{1\over t}\,\rho^{A,V}(t)
\label{IAV}\\
C^{AV}\left(s_0^A,s_0^V\right) &=&\int\limits_{s_0^A}^{s_0^V} dt {1\over
t}\,\tilde\rho(t)\,.
\label{CAV}
\end{eqnarray}

In the above equations,  $s_0^{A,V}$ is the continuum threshold for
duality with asymptotic freedom in the axial and vector channels, {\it i.e.}
for
$t\geq s_0^{A,V}$ the spectral function behaves as predicted by perturbative
QCD.
The spectral function denoted by $\tilde\rho(t)$
represents its  perturbative expression which is the same for the axial and
vector case
for light quarks up to two-loop order.

Since the integrand of $I^A$ is positive definite the following
inequality holds
\cite{DalfovoStringari92} (see \cite{Benmeretal94} and \cite{Bohigasetal79}
for further applications of inequalities and QCD sum-rules)
\begin{equation}
I^A\left(s_0^A\right) \geq L^A\left(s_0^A\right) ={F\over(1-D^2/E)}\,,
\label{LA}
\end{equation}
where $F, D, E$ are combinations of
integrals {\em related to} the finite-energy sum rules (the reason for
this distinction will be seen below)
\begin{equation}
m_n^{A,V}=\int\limits_{4 M^2_{\pi}}^{s_0^{A,V}} dt\, t^n\,\rho^{A,V}(t)
\label{FESR1}
\end{equation}
and in particular
\begin{eqnarray}
F&=&{m_0^2\over m_1}\,,
\label{F}\\
D&=&{m_2\over m_1}-{m_1\over m_0}
\label{D}\\
E&=&{m_3\over m_1}-2{m_2\over m_0}+({m_1\over m_0})^2\,.
\label{E}
\end{eqnarray}
The explicit results for the FESR in the axial and vector channels are
given by
\cite{Bertlmannetal85}
\begin{eqnarray}
{\cal F}_n^{A,V}&=&\int\limits_0^{s_0^{A,V}}t^n
Im\Pi^{A,V}(t)\,dt\label{FESR3}\\
{\cal F}_n^{A,V} &=& {1\over 4 \pi^2} \left[{(s_0^{A,V})^{(n+1)}\over
(n+1)}\left(1+F_{2n+2}(s^{A,V}_0)
\right) + (-1)^n C_{2n+2}<O_{2n+2}>_{A,V}\right]\,,
\label{FESR2}
\end{eqnarray}
where $F_p(s_0)$ are the radiative corrections, which, for three colours
and three flavours are given by
\begin{equation}
F_p(s_0)= {\alpha_S(s_0)\over \pi} + \left[{\alpha_S(s_0)\over \pi}\right]^2
\left(1.641 + {9 \over 2\, p} - {16\over 9} \,\ln \,\ln {s_0 \over
\Lambda_{QCD}^2}\right)
\label{Fp}
\end{equation}
and $C_{2n+2}<O_{2n+2}>_{A,V}$ contain the quark and gluon condensates
\cite{BroadhurstGeneralis85}. Separating out the pion pole contribution from
$ Im\Pi^A(t)$ in (\ref{FESR3}) as required by the DMO sum-rule (\ref{dmo})
gives
\footnote{We are grateful to the referee for bringing this point to our
attention.}
\begin{eqnarray}
{\cal F}_n^A&=&f_\pi^2M_\pi^{2n}+\int\limits_{4M_\pi^2}^{s_0^A}t^n Im\Pi^A(t)dt
\label{pole1}\\
m_n^A&=&{\cal F}_n^A-f_\pi^2M_\pi^{2n}\quad .
\label{mn}
\end{eqnarray}
Of course in the vector channel $m_n^V$ and ${\cal F}_n^V$ are identical.

An upper bound for $I^V\left(s_0^V\right)$ can be found using the following
relation
based again on Schwartz inequalities
\begin{equation}
m_{-2}\geq {FF\over(1-DD^2/EE)}\,,
\end{equation}
where FF, DD and EE are obtained from F, D and E of  (\ref{D}-\ref{E})
replacing
$m_n$ by $m_{n-1}$.
The previous inequality is equivalent to
\begin{equation}
I^V\left(s_0^V\right) \leq  {1\over m_2} \left(m_0 m_1 + \sqrt{m_0^2 m_1^2 -
m_0^3 m_2 + m_{-2} m_2 (m_0 m_2 - m_1^2)}\right)\,.
\end{equation}

Since $m_{-2}$ depends on the extreme infrared properties of QCD
a theoretical prediction for this sum-rule does not exist.
However, by recognizing that the currents
$J^V_\mu(x)=\bar u\gamma_\mu u\,$ and
$J^V_\mu(x)=\bar u\gamma_\mu d\,$ lead to identical correlation functions in
the
$SU(2)$ limit,
an upper bound on $m_{-2}$  can
be found in terms of the hadronic contributions to the
anomalous magnetic moment of the muon. In fact one has the following
series of inequalities \cite{Durand62,Lautrupetal72,Steeleetal91}

\begin{eqnarray}
a_\mu^{had}&=& 4 \alpha^2 \sum_f Q_f^2\int\limits_{4 M^2_{\pi}}^\infty dt\,
K_\mu(t)\,\rho^V(t)
\geq {4 \over 3} \alpha^2\sum_f Q_f^2 M_\mu^2 \int\limits_{4 M^2_{\pi}}^\infty
dt\,
                  {1\over t^2} \,\rho^V(t)\nonumber\\
&\geq& {4 \over 3} \alpha^2\sum_f Q_f^2 M_\mu^2 \int\limits_{4
M^2_{\pi}}^{s_0^V}dt\,
                  {1\over t^2} \,\rho^V(t) = {20 \over 27} \alpha^2 M_\mu^2\,
m_{-2}\,.
\end{eqnarray}
namely
\begin{equation}
m_{-2}\leq {27\over 20}{a_{\mu}^{had}\over \alpha^2 M_{\mu}^2}
\label{m-2}
\end{equation}
and finally

\begin{equation}
I^V\left(s_0^V\right) \leq U^V\left(s_0^V\right) ={1\over m_2} \left(m_0 m_1 +
\sqrt{m_0^2 m_1^2 - m_0^3 m_2 +
{27\over 20}{a_{\mu}^{had}\over \alpha^2 M_{\mu}^2} m_2
(m_0 m_2 - m_1^2)}\right)\,.
\end{equation}
Combining  equations (\ref{I2}), (\ref{IAV}), (\ref{LA}) and (\ref{m-2}) one
has

\begin{equation}
I\geq I^L\left(s_0^A,s_0^V\right)=L^A\left(s_0^V\right)-U^V\left(s_0^V\right)+
C^{AV}\left(s_0^A,s_0^V\right)
\end{equation}
and
\begin{equation}
\tilde\alpha_{_E} \geq \frac{\alpha}{M_\pi
f_\pi^2}I^L\left(s_0^A,s_0^V\right)\,.
\end{equation}

In Figures 1 and 2 the quantity $\alpha I^L\left(s_0^A,s_0^V\right)/(M_\pi
f_\pi^2)$
is plotted as a function of $s_0^A$ and $s_0^V$.
Since only $u$ and $d$ flavours are used in the currents, the value of
$a_{\mu}^{had}$
has been taken equal to $6\times 10^{-8}$ \cite{Steeleetal91,alfamu},
a value which  includes contributions from
the light (u, d) resonances only (strange quarks are a 10\% effect).
As to the condensates,
both standard values \cite{svz} and values from ~\cite{DominguezSola88}
have been used in the FESR.   The figures show that a negative value of
$\tilde\alpha_{_E}$
as found in the resonance saturation approach \cite{Holstein90}
is consistent with our bounds.

The values of the QCD continuum that have been determined from sum-rule
applications
range from $1.5\,{\rm GeV^2}<s_0^V<4.0\,{\rm GeV^2}$ for the vector channel
\cite{svz,DominguezSola88,Bertlmannetal88,Gimenezetal91} and
$1.75\,{\rm GeV^2}<s_0^A<2.5\,{\rm GeV^2}$ in the axial channel
\cite{DominguezSola88,Bertlmannetal88,Gimenezetal91,rry}.
With these bounds on the continuum threshold, Figures 1 and 2 give a lower
bound
on $\tilde\alpha_{_E}$ that lies between $-12.3$ and $-11.3$, reflecting
uncertainties in the QCD condensates and continuum thresholds.

The shaded region of Figure 3 shows the region of $\alpha_{_E}$,
$\langle r_\pi^2\rangle$ parameter space consistent with these lower bounds
on the intrinsic polarizability.
As is evident from the Figure, even the most pessimistic bound
(the lowest diagonal line)
is in disagreement with the largest experimental value of
$\langle r_\pi^2\rangle$ combined with the
chiral perturbation theory or resonance saturation prediction of $\alpha_{_E}$.
In other words, the lower values of
$\alpha_{_E}$ combined with larger values of $\langle r_\pi^2\rangle$ are
incompatible with the QCD determinations of the continuum thresholds
representing the minimum energy necessary for duality.

The present analysis seems to point towards
larger values of $\alpha_{_E}$ and lower values of $\langle r_\pi^2\rangle$.
Further studies of the continuum (duality) thresholds
$s_0^A$ and $s_0^V$, better knowledge of the mean-square pion radius,
and further experimental determinations of $\alpha_{_E}$ would be valuable in
helping to clarify our understanding of QCD.

\vspace{0.5in}

\noindent
{\bf Acknowledgements:} MB and TGS are grateful for the financial support of
the Natural Sciences and Engineering Research Council of Canada (NSERC).
Many thanks to J. Kwan for assistance with the preparation of the figures.

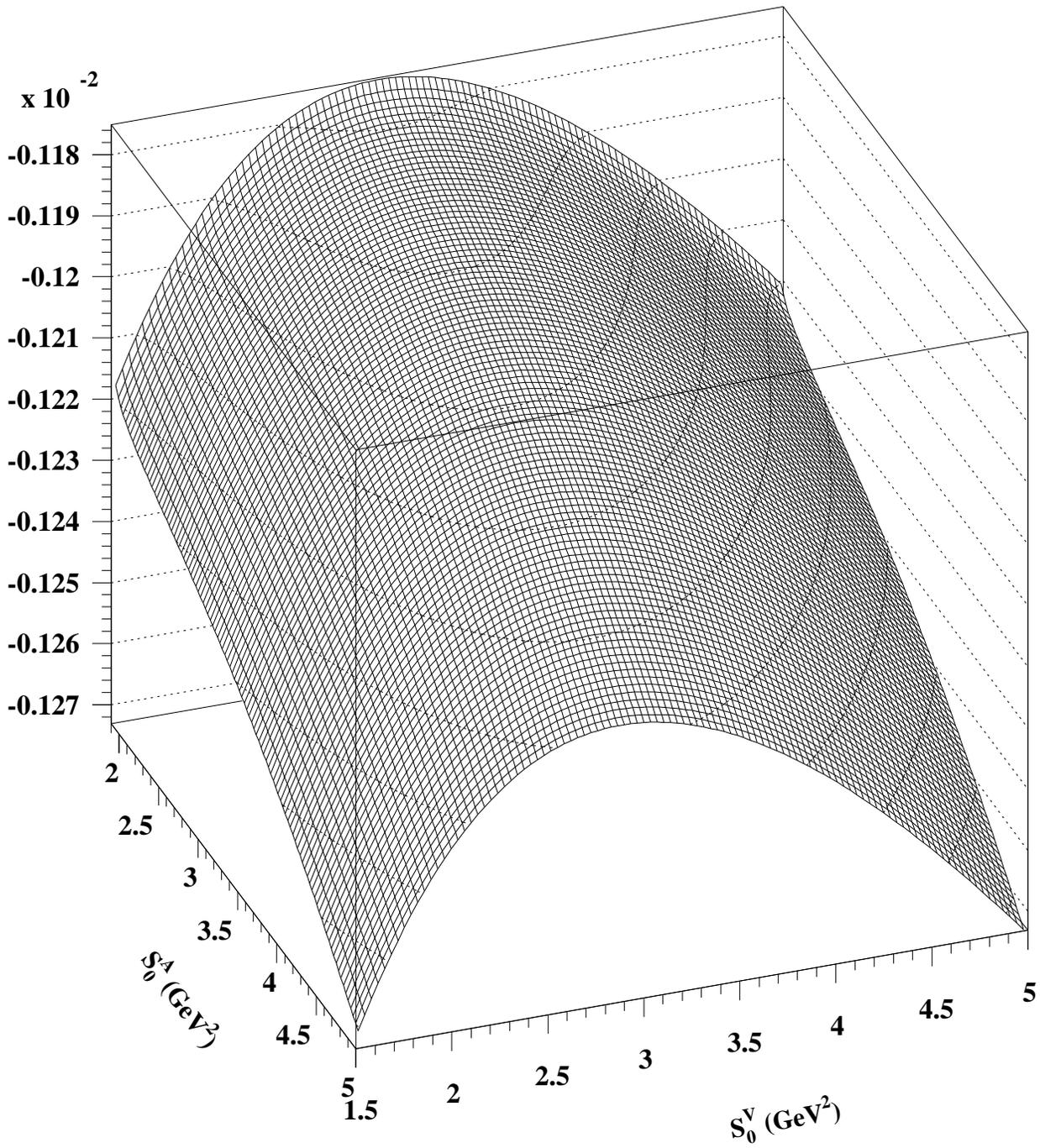
\begin{figure}
\caption{
Plot of $\alpha I^L/(M_\pi f_\pi^2)$ in $fm$ units as a function of
$s_0^A$ and $s_0^V$ for standard values \protect\cite{svz} of the
condensates.}
\label{Figure 1}
\end{figure}

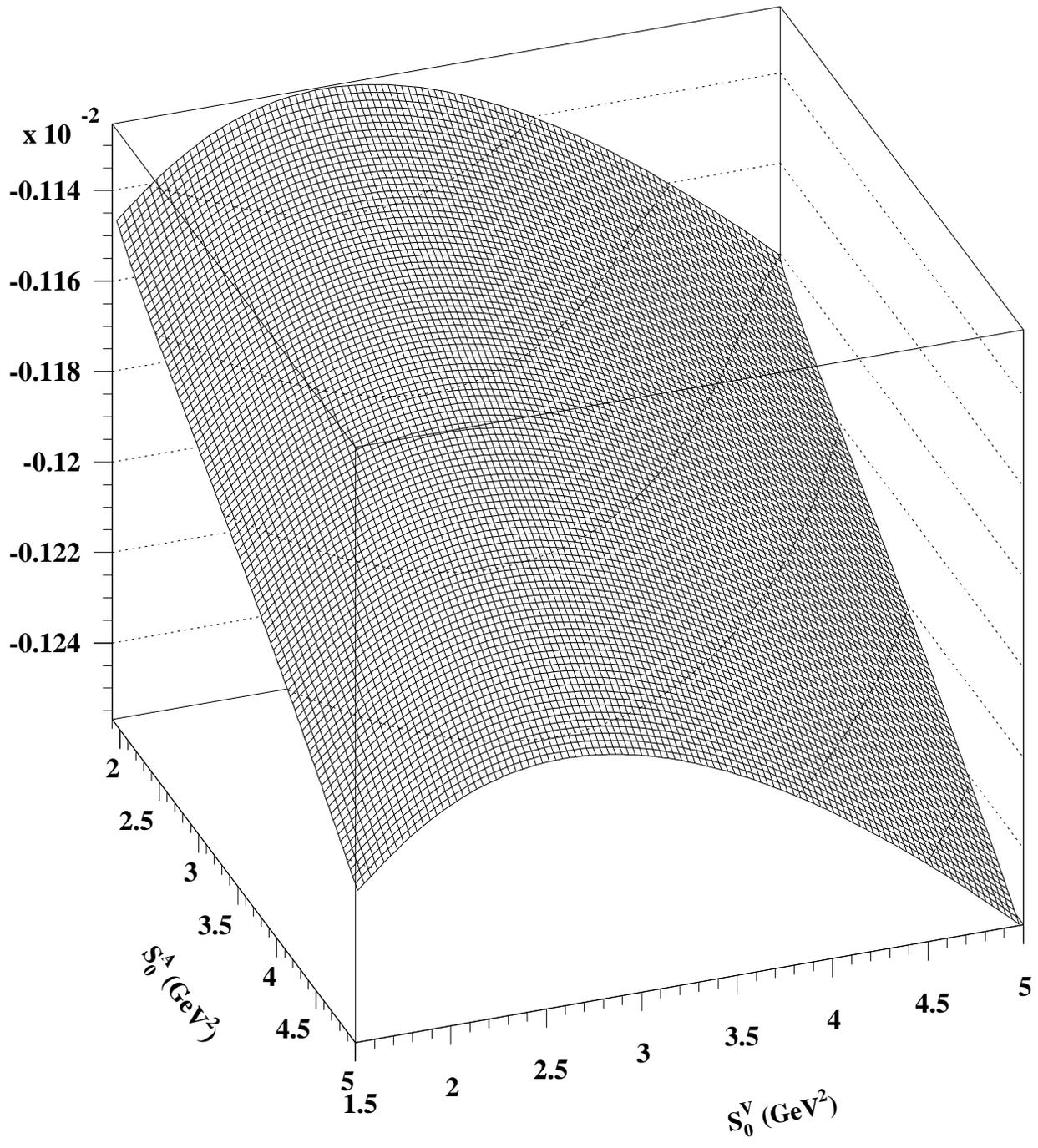
\begin{figure}
\caption{Same as Figure 1 except for using values of the condensates
as determined in \protect\cite{DominguezSola88}.}
\label{Figure 2}
\end{figure}

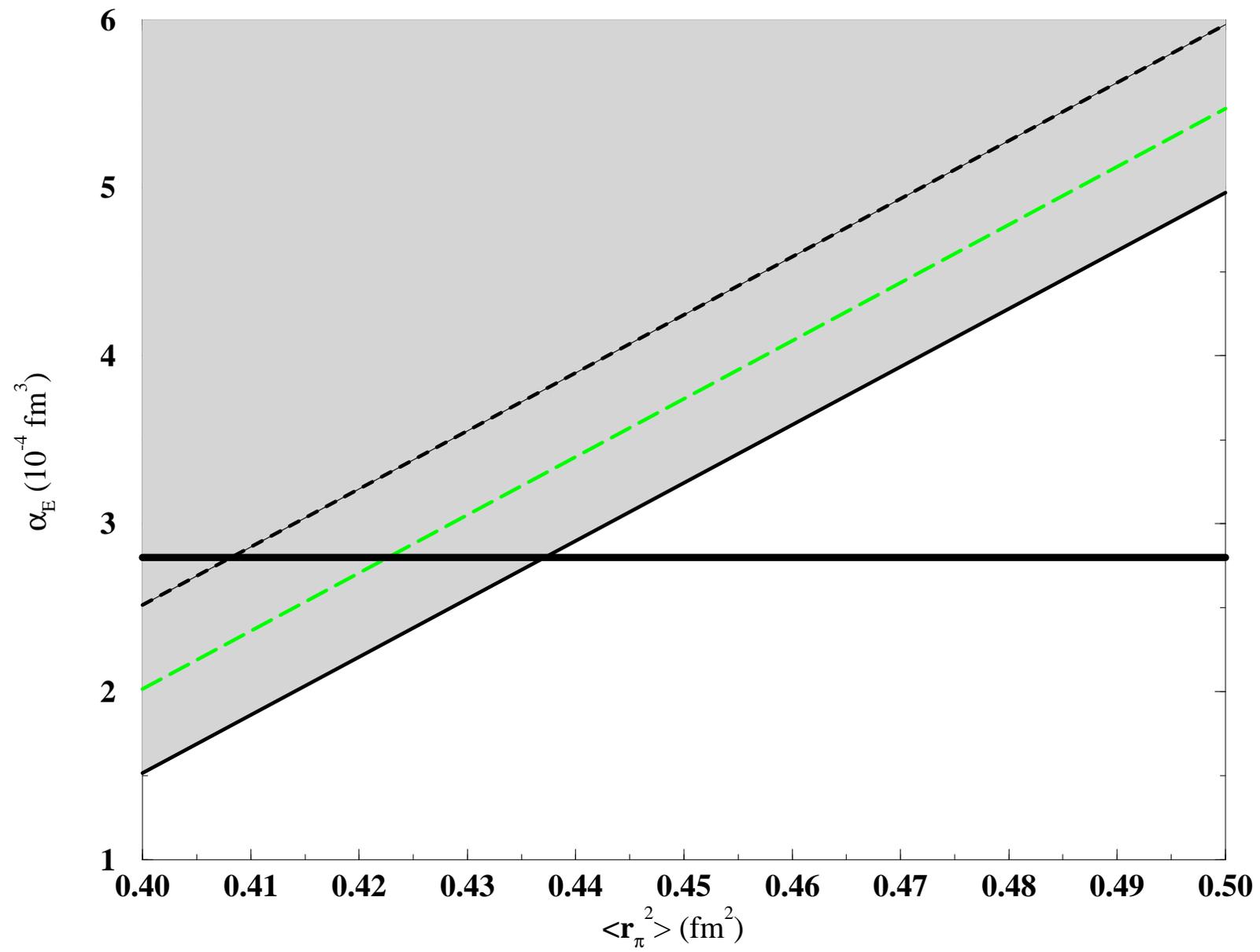
\begin{figure}
\caption{
Shaded region represents the $\alpha_{_E}$, $\langle r_\pi^2\rangle$
parameter space consistent with our QCD sum-rule bound on
$\tilde\alpha_{_E}$.  Diagonal
lines represent possible borders of the parameter space for various choices
of the condensates and continuum threshold.  The horizontal line represents
the chiral perturbation theory prediction of $\alpha_{_E}.$ }
\label{Figure 3}
\end{figure}

\end{document}